\documentclass[a4paper,aps,prb,reprint,showpacs,twocolumns,superscriptaddress]{revtex4-1}
\usepackage{amsmath,amssymb,mathrsfs}
\usepackage[pdftex]{graphicx}
\usepackage{color}

\newcommand{\ket}[1]{\left|#1\right\rangle}
\newcommand{\bra}[1]{\left\langle#1\right|}

\newcommand{\ii}{\text{i}}
\newcommand{\dd}{\text{d}}

\allowdisplaybreaks[4]

\begin{document}

\title{Observation of spin-charge separation and boundary bound states via the local density of states}
\author{Benedikt Schoenauer}
\affiliation{Institute for Theoretical Physics, Center for Extreme Matter and Emergent Phenomena, Utrecht University, Princetonplein 5, 3584 CE Utrecht, The Netherlands}
\email{b.m.schonauer@uu.nl, d.schuricht@uu.nl}
\author{Peter Schmitteckert}
\affiliation{Lehrstuhl f{\"u}r Theoretische Physik I, Physikalisches Institut, Am Hubland, Universit{\"a}t  W{\"u}rzburg, 97074 W{\"u}rzburg, Germany}
\author{Dirk Schuricht}
\affiliation{Institute for Theoretical Physics, Center for Extreme Matter and Emergent Phenomena, Utrecht University, Princetonplein 5, 3584 CE Utrecht, The Netherlands}
\date{\today}
\pagestyle{plain}

\begin{abstract}
We numerically calculate the local density of states (LDOS) of a one-dimensional Mott insulator with open boundaries, which is modelled microscopically by a (extended) Hubbard chain at half filling. In the Fourier transform of the LDOS we identify several dispersing features corresponding to propagating charge and spin degrees of freedom, thus providing a visualisation of the spin-charge separation in the system. We also consider the effect of an additional boundary potential, which, if sufficiently strong, leads to the formation of a boundary bound state which is clearly visible in the LDOS as a non-dispersing feature inside the Mott gap.
\end{abstract}
\pacs{68.37.Ef, 71.10.Pm, 71.10.Fd, 73.20.At}
\maketitle


\section{Introduction}\label{sec:intro}
One-dimensional systems remain a fascinating field in condensed-matter physics since they constitute prime examples for the breakdown of Fermi-liquid theory, which has to be replaced by the Luttinger-liquid paradigm.\cite{Giamarchi04} Arguably the most dramatic consequence of this is the absence of electron-like quasiparticles, manifesting itself in the separation of spin and charge degrees of freedom visible for example in angle-resolved photoemission,\cite{Kim-96} transport,\cite{Bockrath-99} scanning tunneling spectroscopy\cite{Lee-04} or resonant inelastic X-ray scattering\cite{Schlappa-12} experiments as well as analytical\cite{BaresBlatter90} and numerical studies of several one-dimensional models.\cite{Jagla-93}

The spectral properties of one-dimensional electron systems have been intensively investigated in the past. These works considered the gapless Luttinger liquid,\cite{Giamarchi04,MedenSchoenhammer92} gapped systems like Mott insulators or charge-density wave states,\cite{Preuss-94} Luttinger liquids with impurities,\cite{FabrizioGogolin95} corrections to the Luttinger model due to non-linear dispersions\cite{ImambekovGlazman09} or the momentum dependence of the two-particle interaction,\cite{MarkhofMeden16} as well as additional phonon degrees of freedom.\cite{Meden-94} These investigations uncovered universal power-law behaviour at low energies as well as deviations thereof, spin-charge separation visible in the propagation modes, and signatures of these features in various experimental probes.

In this article we consider another situation, namely the microscopic study of the boundary effects on one-dimensional Mott insulators. Specifically we numerically study the local density of states (LDOS) of one-dimensional Hubbard models with open, ie, hard-wall, boundary conditions, where the system is at half filling and thus in its Mott phase. A previous field-theoretical analysis\cite{SEJF08} has shown that the Fourier transform of the LDOS\cite{Kivelson-03} exhibits clear signatures of propagating spin and charge degrees of freedom, thus providing a way to detect spin-charge separation. Furthermore, an additional boundary potential may lead to the formation of a boundary bound state, which manifests itself as a non-dispersing feature in the LDOS. The aim of our work is to calculate the Fourier transform of the LDOS directly in the microscopic lattice model using  a multi-target\cite{Schmitteckert04,BraunSchmitteckert14} variant of the density matrix renormalisation group (DMRG) method \cite{White92} employing an expansion in Chebyshev polynomials. We find our numerical results to be fully consistent with the analytical predictions both qualitatively, ie, concerning the number of dispersion modes and their basic properties, as well as quantitatively with respect to the numerical values of the effective parameters like the Mott gap and spin and charge velocities as compared to the exact results obtained from the Bethe ansatz.\cite{EsslerFrahmGoehmannKluemperKorepin05} Thus our work provides a microscopic calculation of the Fourier transform of the LDOS in a gapped, strongly correlated electron system, showing spin-charge separation as well as the formation of a boundary bound state. 

This paper is organised as follows: In Sec.~\ref{sec:model} we present the microscopic models to be analysed and discuss the basic setup. In Sec.~\ref{sec:method} we give a brief summary of the numerical method we employ to calculate the single-particle Green function. Our results for the LDOS of the Mott insulators with open boundary conditions are discussed in Sec.~\ref{sec:ldos}. In Sec.~\ref{sec:bbs} we study the effect of a boundary potential on the LDOS, in particular we analyse the properties of the boundary bound state existing for sufficiently strong boundary potentials. In Sec.~\ref{sec:conclusion} we summarise our results. 

\section{Model}\label{sec:model}
In this work we analyse the LDOS of the one-dimensional Hubbard model~\cite{EsslerFrahmGoehmannKluemperKorepin05} at half filling. The Hamiltonian is given by
\begin{align}
H=&-t \sum_{\sigma,j=0}^{L-2}\bigl( c_{j,\sigma}^{\dagger}c_{j+1,\sigma} + c_{j+1,\sigma}^{\dagger}c_{j,\sigma}\bigr)\label{eq:Hubbard}\\*
 &+U\sum_{j=0}^{L-1} \left(n_{j,\uparrow}-\frac{1}{2}\right)\left(n_{j,\downarrow}-\frac{1}{2}\right)\nonumber,
\end{align}
where $c_{j,\sigma}$ and $c_{j,\sigma}^\dagger$ denote the annihilation and creation operators for electrons with spin $\sigma=\uparrow,\downarrow$ at lattice site $j$ and $n_{j,\sigma}=c_{j,\sigma}^{\dagger}c_{j,\sigma}$ the corresponding density operators. The parameters $t$ and $U>0$ describe the hopping and repulsive on-site interaction respectively. Furthermore we consider a chain with $L$ sites and open boundary conditions. Since the system is assumed to be at half filling, the Fermi momentum is given by $k_\text{F}=\pi/2$. 

As is well known,\cite{Giamarchi04,EsslerFrahmGoehmannKluemperKorepin05} in the Hubbard model at half filling, ie, when there are $L$ electrons in the system, the repulsive interaction opens a gap in the charge sector and the system becomes a Mott insulator. Using bosonisation the low-energy behaviour of the system is described by the massive Thirring model;\cite{EsslerKonik05} the LDOS of which in the presence of boundaries has been analysed in Refs.~\onlinecite{SEJF08}. The main objective of our article is the comparison of the LDOS of the Hubbard model \eqref{eq:Hubbard} with the field-theoretical results obtained in the Thirring model. Hereby the effective parameters in the field theory, ie, the mass gap and velocities, can be obtained from the exact Bethe-ansatz solution of the Hubbard model. This allows us to choose the microscopic parameters such that the expected features of the Fourier transformed LDOS can be easily resolved in the numerical results.

In addition to the standard Hubbard model \eqref{eq:Hubbard} we also consider its extension including a nearest-neighbour interaction $V$, ie, the Hamiltonian is given by\cite{Voit92}
\begin{equation}
H_\text{ext}=H+V\sum_{j=0}^{L-2}\left(n_{j}-1\right)\left(n_{j+1}-1\right)
\label{eq:ext}
\end{equation}
where $n_j=n_{j,\uparrow}+n_{j,\downarrow}$ is the total electron density. The low-energy regime of the extended Hubbard model \eqref{eq:ext} is still described~\cite{EsslerKonik05} by the massive Thirring model. However, since \eqref{eq:ext} is no longer integrable, the explicit relation between the microscopic parameters $t$, $U$ and $V$ and the field-theory ones is not known. Thus the investigation of the phase diagram of the extended Hubbard model at half filling had to be performed by numerical means.\cite{Clay-99} Using these results we choose the microscopic parameters such that the system is well inside the Mott-insulating phase with an energy gap $\Delta\approx \mathcal{O}(t)$ so that we are able to clearly resolve the interesting features in our numerical results.

\section{Green function}\label{sec:method}
In order to determine the LDOS we calculate the retarded Green function in frequency space using an expansion of the occurring resolvent in Chebyshev polynomials.\cite{BraunSchmitteckert14} An alternative numerical approach consists in the expansion of the Lehmann representation of the spectral function in Chebyshev polynomials, the kernel polynomial method (KPM), see Refs.~\onlinecite{Weisse-06}. In contrast we specifically evaluate the complete (real and imaginary part) Green functions
\begin{equation}
G^\text{R}(\omega,x)=G^+(\omega,x)-G^-(\omega,x)
\label{eq:GR}
\end{equation}
with
\begin{eqnarray}
G^+(\omega,x)&=&\bra{\Psi_0}c_{j,\sigma}\frac{1}{E_0-H+\omega+\ii\eta}c_{j,\sigma}^\dagger\ket{\Psi_0},\label{eq:G+}\\
G^-(\omega,x)&=&\bra{\Psi_0}c_{j,\sigma}^\dagger\frac{1}{E_0-H-\omega-\ii\eta}c_{j,\sigma}\ket{\Psi_0}.\label{eq:G-}
\end{eqnarray}
Here $\ket{\Psi_0}$ denotes the ground state of the system with energy $E_0$. Note that since we are interested in the LDOS we have already taken the electron creation and annihilation operators to be at the same site $x=ja_0$ with $a_0$ denoting the lattice spacing. Furthermore, since the systems we consider possess spin-rotation invariance we have suppressed the formal spin dependence of the Green functions.

In Eqs.~\eqref{eq:G+} and~\eqref{eq:G-} we have included the convergence factor $\eta$, which in the continuum limit should be taken as $\eta\rightarrow 0^{+}$. In the numerical evaluations it has to be larger than the finite level splitting brought about by the finite system size. At the same time $\eta$ has to be smaller than any physically relevant energy scale in order to resolve the relevant features of the spectrum. To attain a small value of $\eta$ we employ a Chebyshev polynomial expansion approach for the resolvents in~\eqref{eq:G+} and~\eqref{eq:G-}. More details on this approach can be found in Refs.~\onlinecite{Schmitteckert10,BraunSchmitteckert14}.

The applied Chebyshev expansion is based on the representation of the functions
\begin{equation}
f^\pm(\omega,z)=\frac{1}{\pm \omega-z}
\end{equation}
in terms of Chebyshev polynomials
\begin{equation}
f^{\pm}(\omega,z)=\sum_{n=0}^{\infty} \alpha_n^{\pm}(\omega) T_n(z),\quad -1\le z\le 1.
\label{eq:fexpansion}
\end{equation}
The expansion coefficients are given by
\begin{align}
\alpha_{n}^{\pm}(\omega) &= \frac{2}{\pi (1+\delta_{n,0})}\int_{-1}^{1}\dd z\,\frac{T_n(z)}{\sqrt{1-z^2}}\frac{1}{\pm \omega-z}\\\nonumber
&=\frac{2-\delta_{n,0}}{(\pm \omega)^{n+1}\left(1+\sqrt{\omega^2}\frac{\sqrt{\omega^2-1}}{\omega^2}\right)^{n}\sqrt{1-\omega^{-2}}},
\end{align}
where $\alpha_n^{\pm}(\omega)\equiv\alpha_n^{\pm}(\omega+\text{i}\eta)$ is a function of the artificial broadening $\eta$ which would theoretically allow arbitrarily small $\eta$. The Chebyshev polynomials $T_n(z)$ are defined by their recursion relation
\begin{align}
T_0(z)&=1,\\
T_1(z)&=z,\\
T_{n+1}(z)&=2zT_n(z)-T_{n-1}(z),\quad n\geq2,
\end{align}
and fulfil
\begin{align}
\int_{-1}^{1} \frac{\dd z}{\sqrt{1-z^2}}\,T_n(z) T_m(z)=\frac{\pi}{2}\delta_{n,m}(1+\delta_{n,0})
\end{align}
as well as
\begin{align}
T_{2n}(z)&=2T_n(z)^2-T_0(z),\label{eq:Chaddtheo1}\\
T_{2n-1}(z)&=2T_{n-1}(z)\,T_n(z)-T_1(z).\label{eq:Chaddtheo2}
\end{align}
In order to apply the expansion \eqref{eq:fexpansion}, which is only valid for $|z|\le 1$, to the resolvents appearing in the Green functions, we first have to rescale the energies. To this end we run initial DMRG calculations to determine the ground-state energy $E_0$ as well as the smallest and the largest energies of the system with $L\pm 1$ electrons. This allows us to find the scaling factor $a$ and shift $b$ such that the operator 
\begin{equation}
a(H-E_0)-b 
\end{equation}
has a spectrum between $-1$ and $1$ in the sectors with $L\pm 1$ particles. Then the Green function \eqref{eq:G+} can be expressed as
\begin{equation}
G^+(\omega,x)=a\sum_{n=0}^\infty \alpha_n^+[a(\omega+\ii\eta)-b]\,\mu_n^+(x),
\label{eq:G+expansion}
\end{equation}
where the Chebyshev moments
\begin{equation}
\mu_n^+(x)=\bra{\Psi_0}c_{j,\sigma}\,T_n[a(H-E_0)-b]\,c_{j,\sigma}^\dagger\ket{\Psi_0}
\end{equation}
(recall $x=ja_0$) can be evaluated recursively via 
\begin{align}
\mu_{n}^{+} (x)= \bra{\Psi_0}c_{\sigma}(x)\ket{\Phi_{n}^{+}}
\end{align}
with the recursion relations
\begin{align}
\vert \Phi_{0}^{+} \rangle &= c^{\dagger}_{\sigma}(x)\vert \Psi_0\rangle,\label{eq:iteration1}\\
\vert \Phi_{1}^{+} \rangle &= [a(H-E_0)-b]\vert \Phi_0^{+}\rangle,\label{eq:iteration2}\\
\vert \Phi_{n+1}^{+} \rangle &= 2[a(H-E_0)-b]\vert \Phi_{n}^{+}\rangle -\vert \Phi_{n-1}^{+}\rangle.\label{eq:iteration3}
\end{align}
Similarly, for the Green function \eqref{eq:G-} we obtain the expansion 
\begin{equation}
G^-(\omega,x)=a\sum_{n=0}^\infty \alpha_n^-[a(\omega+\ii\eta)+b]\,\mu_n^-(x),
\label{eq:G-expansion}
\end{equation}
where 
\begin{equation}
\mu_n^-(x)=\bra{\Psi_0}c_{j,\sigma}^\dagger\,T_n[a(H-E_0)-b]\,c_{j,\sigma}\ket{\Psi_0}.
\end{equation}
In the numerical evaluations the sums appearing in \eqref{eq:G+expansion} and \eqref{eq:G-expansion} are truncated at $N/2$. The moments $\mu_n^\pm$ are calculated iteratively from \eqref{eq:iteration1}--\eqref{eq:iteration3} using DMRG. During the DMRG finite-lattice sweeps we determine each state $\ket{\Phi_0^\pm}, \ldots, \ket{\Phi_{N/2}^\pm}$ and include it into a modified density matrix. By performing a singular-value decomposition of this modified density matrix we ensure that all the states $\ket{\Phi_0^\pm}, \ldots, \ket{\Phi_{N/2}^\pm}$ are part of the Hilbert space after the DMRG truncation. The moments for $n=N/2+1,\ldots,N$ are then obtained employing \eqref{eq:Chaddtheo1} and \eqref{eq:Chaddtheo2} as $\mu_{2n}^\pm=2\bra{\Phi_n^\pm}\Phi_n^\pm\rangle-\bra{\Phi_0^\pm}\Phi_0^\pm\rangle$ and $\mu_{2n-1}^\pm=2\bra{\Phi_{n-1}^\pm}\Phi_n^\pm\rangle-\bra{\Phi_0^\pm}\Phi_1^\pm\rangle$.

Finally, we note that the Chebyshev moments $\mu_n^\pm$ are typically strongly oscillating with respect to the index $n$. Therefore, the final results oscillate slightly when changing the value of $N$. On the other hand, we find small oscillating parts in the spectral function if we choose $N$ too small. Both effects can be avoided by implementing a smoothing window for the last $N_\text{S}$ moments. Throughout this article we use a $\cos^2$-filter for the last $N_\text{S}=N/5$ moments. This way one can obtain a good approximation for the spectral function using a smaller number of moments $N$. Previously it was observed\cite{BraunSchmitteckert14} that the number of required Chebyshev moments sufficient to approximate the Green function is inversely proportional to the width of the spectrum $a$ and the desired artificial broadening $\eta$, ie, $N\simeq (a\eta)^{-1}$. Throughout this work use $N\geq 1000$ Chebyshev moments for the series expansion of the Green function. Furthermore, $\eta$ is chosen such that the resulting curves become smooth and artificial features are suppressed. 

\section{LDOS}\label{sec:ldos}
The LDOS is obtained from the retarded Green function \eqref{eq:GR} in the usual way. As was noted by Kivelson et al.\cite{Kivelson-03} in the study of Luttinger liquids with boundaries, it is useful to consider the Fourier transform of the LDOS, as physical properties like the dispersions of propagating quasiparticles can be more easily identified. Since we consider a finite chain of length $L$ we analyse 
\begin{equation}
N(\omega,Q)=-\frac{1}{\pi}\sqrt{\frac{2}{L+1}}\sum_{j=0}^{L-1}\, \text{Im}\, G^\text{R}(\omega,x)\,\sin[Q(j+1)],
\label{eq:NQ}
\end{equation}
where the momenta $Q$ take the values $Q=\pi k/(L+1)$, $k=1,\ldots, L$. We note that the LDOS is directly related to the tunneling current measured in scanning tunneling microscopy experiments, thus its Fourier transform \eqref{eq:NQ} is experimentally accessible. In the following we focus on the LDOS for positive energies;  the LDOS for negative energies can be analysed analogously. 

The LDOS of the low-energy effective field theory of the Hubbard models \eqref{eq:Hubbard} and \eqref{eq:ext} has been analysed\cite{footnote1} in Refs.~\onlinecite{SEJF08}. In the field-theoretical description the momentum regimes $Q\approx 0$ and $Q\approx\pm 2k_\text{F}=\pi$ are treated separately. For small momenta $Q\approx 0$ the main features of the Fourier transform \eqref{eq:NQ} are a strong divergence at $Q=0$ as well as a propagating excitation in the gapped charge sector above the Mott gap. In contrast, the behaviour at momenta $Q\approx 2k_\text{F}$ shows a divergence at $Q=2k_\text{F}$, a propagating excitation in the charge sector as well as a linearly dispersing excitation in the gapless spin sector. Furthermore there exists a critical momentum above which a second linearly dispersing mode becomes visible. In addition, it was shown that certain boundary conditions lead to the formation of boundary bound states which manifest themselves as non-propagating features in the LDOS. 

The main aim of our article is the calculation of the Fourier transform of the LDOS \eqref{eq:NQ} in the microscopic models \eqref{eq:Hubbard} and \eqref{eq:ext} and its comparison to the field-theoretical predictions.\cite{SEJF08} We start with the standard Hubbard chain \eqref{eq:Hubbard} before considering the extended version \eqref{eq:ext}. In Sec.~\ref{sec:bbs} we then analyse the effect of additional boundary potentials which give rise to the existence of boundary bound states. 

\subsection{Standard Hubbard model}\label{sec:std}
\begin{figure}[t]
\includegraphics[width=0.495\textwidth]{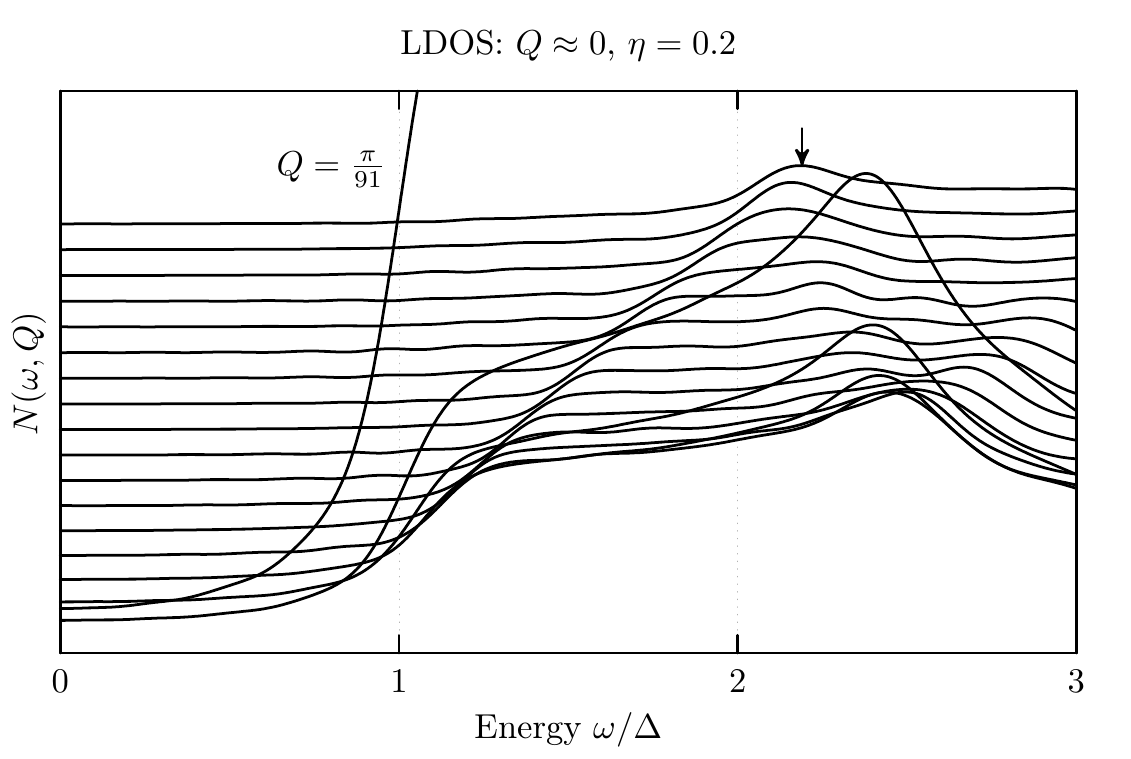}
\caption{Fourier transform of the LDOS, $N(\omega,Q)$, for interaction $u=U/4=1.125$ (recall $t=1$), $L=90$ lattice sites, broadening $\eta=0.2$ and momenta $Q=\pi/91, 2\pi/91,\ldots, 18\pi/91$ (from bottom to top). The curves are constant $Q$-scans that have been offset along the y-axis by a constant with respect to one another. $N(\omega,Q)$ is dominated by a strong peak at $Q=\pi/91\approx0$ which is only partially displayed in the figure in order to improve visibility for the other cuts. We clearly observe the Mott gap $\Delta$ as well as a dispersing feature indicated by the arrow. This feature corresponds to propagating charge excitations, it follows the dispersion relation $E_\text{c}(Q)$ given in \eqref{eq:holon} with $v_\text{c}\simeq 2.67$ obtained from \eqref{eq:BAvc}.}
\label{fig:std_q0}
\end{figure}

\begin{figure}[t]
\includegraphics[width=0.495\textwidth]{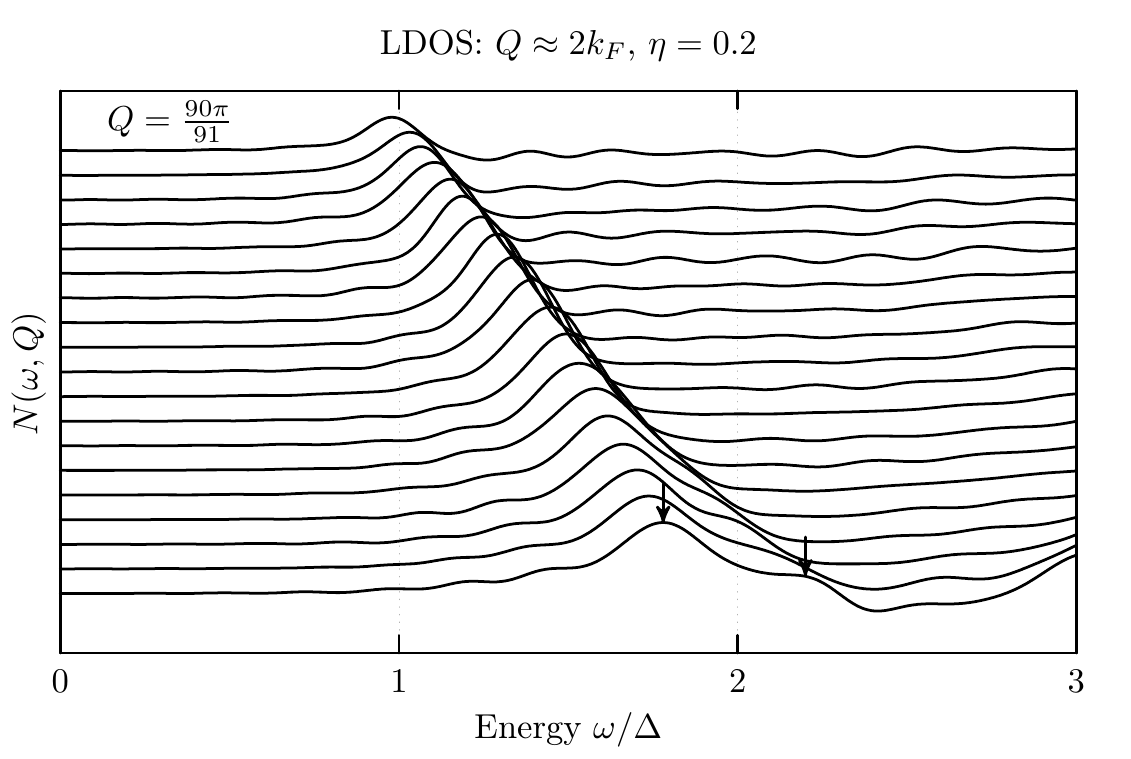}
\caption{$N(\omega,2 k_\text{F}-q)$ for momenta in the vicinity of $Q=2 k_\text{F}=\pi$ with $q=2k_\text{F}-Q=\pi/91,2\pi/91,\ldots, 19\pi/L$ (from top to bottom). All other parameters are as in Fig.~\ref{fig:std_q0}. The curves are constant $q$-scans that have been offset along the y-axis by a constant with respect to one another. We observe two dispersing features (indicated by the arrows) at $E_\text{c}(q)$ and $E_\text{s}(q)$ originating from propagating charge and spin excitations respectively.}
\label{fig:std_q2kf}
\end{figure}

We first consider the Fourier transform of the LDOS \eqref{eq:NQ} in the standard Hubbard model \eqref{eq:Hubbard}. The results in the vicinity of $Q=0$ and $Q=2k_\text{F}=\pi$ are shown in Figs.~\ref{fig:std_q0} and~\ref{fig:std_q2kf} respectively, where we have chosen a repulsive interaction of $U=4.5\, t$ corresponding to the dimensionless Hubbard parameter $u=U/(4t)=1.125$ and $L=90$ lattice sites. Throughout our article we use the hopping parameter $t=1$ as our unit of energy. 

As is well known, the Hubbard model \eqref{eq:Hubbard} is exactly solvable by Bethe ansatz.~\cite{EsslerFrahmGoehmannKluemperKorepin05} In particular, the velocities of the spin and charge excitations $v_\text{s}$ and $v_\text{c}$ as well as the Mott gap $\Delta$ can be determined analytically; the results in the thermodynamic limit read
\begin{align}
&\Delta= -2+2u+2\int_{0}^{\infty}\frac{\dd\omega}{\omega}\,\frac{J_1(\omega)\text{e}^{-u\omega}}{\cosh\left(u\omega\right)},\label{eq:Mottgap}\\
&v_\text{c} =\frac{2}{1-\xi_{0,0}(u)}\,\sqrt{u-1+\xi_{-1,1}(u)} \sqrt{1-\xi_{1,1}(u)},\label{eq:BAvc}\\
&v_\text{s}=\frac{2I_1\left(\frac{\pi}{2u}\right)}{I_0\left(\frac{\pi}{2u}\right)},\label{eq:BAvs}\\
&\xi_{m,n}(u)=2\int_0^{\infty}\frac{\dd\omega\,\omega^m J_n(\omega)}{1+\exp(2\omega u)},
\end{align}
where $J_{n}(z)$ and $I_n(z)$ denote the Bessel functions and modified Bessel functions of the first kind respectively. Our chosen parameters for the microscopic system correspond to\cite{footnote1} $v_\text{c} > v_\text{s}$. 

In Fig.~\ref{fig:std_q0} we plot $N(\omega,Q)$ in the vicinity of small momenta $Q\approx 0$. All features (except for a strong peak at $Q=\pi/91$) appear at energies $\omega\ge\Delta$, clearly showing that the system is in a gapped phase. The observed energy gap $\Delta$ agrees perfectly with the value $\Delta(u=1.125)\simeq 0.83$ obtained from the Bethe ansatz \eqref{eq:Mottgap} in the thermodynamic limit. This suggests that the length of our chain is long enough to avoid significant finite-size effects in our results. The Fourier transform of the LDOS for small momenta is dominated by a global maximum at $Q=\pi/91\approx 0$. This peak is attributed to a spin-density wave pinned at the boundary, it is also well visible in the field-theoretical results.\cite{SEJF08} At low energies above the energy gap we further observe a dispersing feature indicated by the arrow. This again agrees well with the results from the field theory that predict a gapped, dispersing charge excitation with dispersion relation
\begin{align}
E_\mathrm{c}(q)=\sqrt{\left(\frac{v_\text{c}q}{2}\right)^2+\Delta^2},
\label{eq:holon}
\end{align}
where $q=Q$ and $v_\text{c}$ is the velocity of the charge excitations. The Bethe-ansatz solution \eqref{eq:BAvc} gives the value $v_\text{c}(u=1.125)\simeq 2.67$, which is in excellent agreement with the velocity observed in  the plot. The physical origin of this dispersing feature is the decay of the electronic excitation into gapped charge and gapless spin excitations. In the process giving rise to \eqref{eq:holon} the external momentum $q$ is taken by the charge excitation propagating through the system and eventually getting reflected at the boundary, while the spin excitation does not propagate and thus possesses zero momentum. The appearance of $v_\text{c}/2$ in \eqref{eq:holon} originates from the fact that the charge excitation has to propagate to the boundary and back, thus covering the distance $2x$. In addition, we find a second dispersing feature that seemingly follows the same dispersion relation albeit with a different value for the gap $\Delta_2 \simeq 5\Delta/2$. This feature is not contained in the field-theory description of the LDOS, which solely focuses on the low-energy regime. Furthermore, the field theory makes predictions about the power-law decay of $N(\omega,Q)$ at $Q=0$ which, however, cannot be resolved in our numerical data. For the observation of such features we would require a significantly higher resolution, both in energy and momentum. This can in turn only be achieved by turning to a significantly larger system size and a higher amount of calculated Chebyshev moments.

We now turn our attention to momenta in the vicinity of $Q=2 k_\text{F}=\pi$. We first note that features in this momentum regime originate from umklapp processes coupling left- and right-moving modes which are absent in translationally invariant systems and thus constitute a particularly clean way to investigate the boundary effects. In Fig.~\ref{fig:std_q2kf} we again observe the existence of the Mott gap as well as two dispersing features at $E_\text{c}(q)$ as defined in \eqref{eq:holon} and 
\begin{align}
E_\text{s}(q)=\frac{v_\text{s} \vert q \vert}{2}+\Delta,
\label{eq:spinon}
\end{align}
both indicated by the arrows. The spin velocity observed in the plot is in excellent agreement with the Bethe-ansatz result \eqref{eq:BAvs} giving $v_\text{s}(u=1.125)\simeq 1.14$. While the feature adhering to \eqref{eq:holon} is again due to a propagating charge excitation, the feature following \eqref{eq:spinon} originates from the propagation of spin excitations with the charge excitation possessing zero momentum. Furthermore, we note that in contrast to the field-theoretical prediction we observe only one linearly dispersing mode. In order to understand this we recall that the two linearly dispersing modes are energetically separated by\cite{SEJF08} $\Delta[1-\sqrt{1-(v_\text{s}/v_\text{c})^2}]\approx 0.1\Delta\approx 0.08$, where in the last step we have put in the parameters used in Fig.~\ref{fig:std_q2kf}. On the other hand, our resolution in energy is limited by finite-size effects to about  $\sim 2\pi/L$, implying that for the treatable system sizes the two linearly dispersing features cannot be separated. 

\begin{figure}[t]
\includegraphics[width=0.495\textwidth]{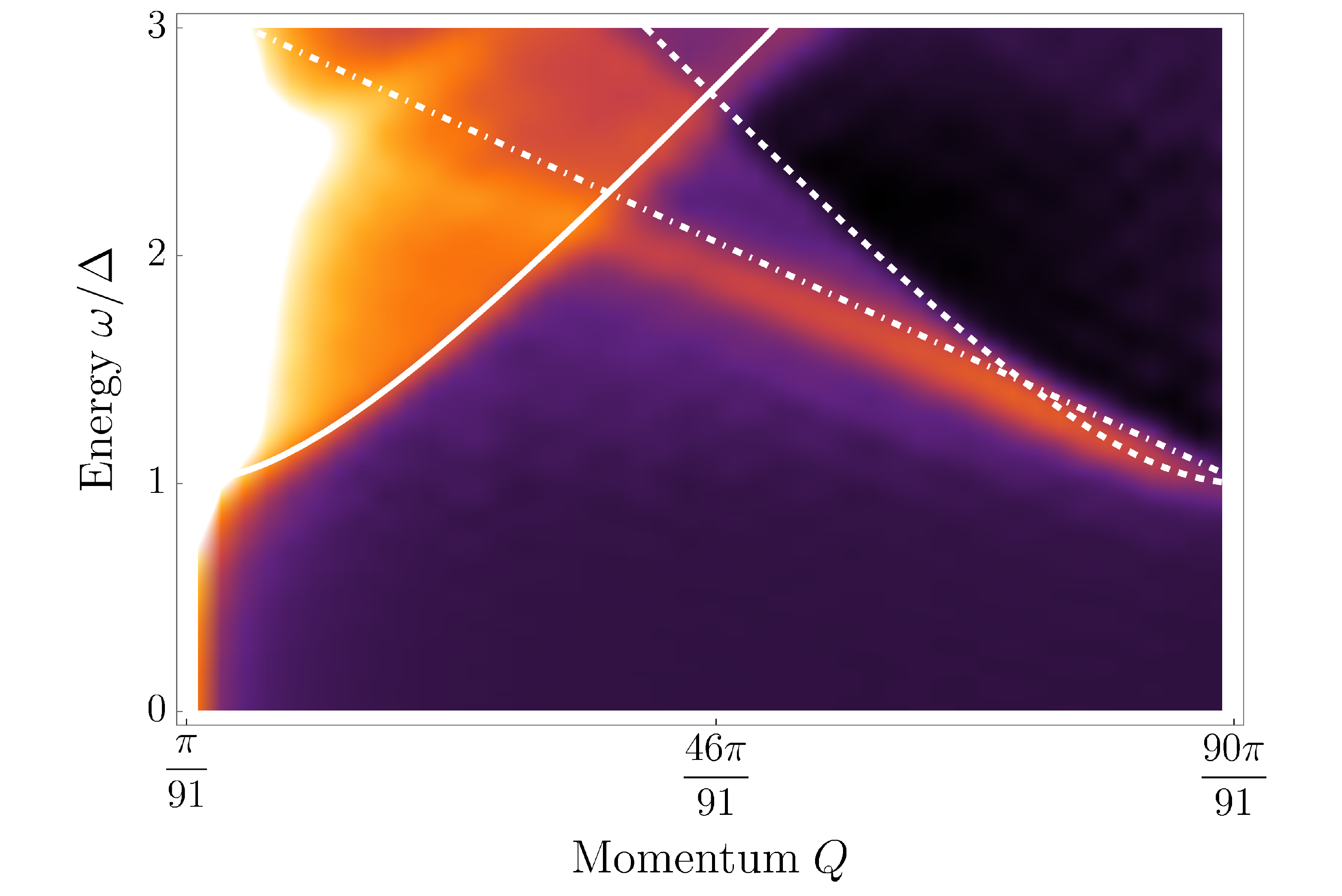}
\caption{Contour plot of the LDOS $N(\omega,Q)$ for the parameters of Fig.~\ref{fig:std_q0}. The dominant, white peak at $Q\approx 0$ is due to the spin-density wave pinned at the boundary. The solid and dashed lines indicate the holon dispersion \eqref{eq:holon} around $Q=0$ and $Q=2k_\text{F}$ respectively, the dashed-dotted  line represents the spinon dispersion \eqref{eq:spinon} around $Q=2k_\text{F}$. The parameters $\Delta$, $v_\text{c}$ and $v_\text{s}$ used in the plot were obtained from the Bethe ansatz for the bulk system \eqref{eq:Mottgap}--\eqref{eq:BAvs}, ie, there is no free fitting parameter.}
\label{fig:modes}
\end{figure}
To summarise our results, in Fig.~\ref{fig:modes} we show a contour plot of the LDOS. For comparison we plot the holon dispersion \eqref{eq:holon} around $Q=0$ and $Q=2k_\text{F}$ as well as the spinon dispersion \eqref{eq:spinon} around $Q=2k_\text{F}$, for which we used the parameters $\Delta$, $v_\text{c}$ and $v_\text{s}$ obtained from the Bethe ansatz for the bulk system. In particular, we stress that there is no fitting parameter.  In conclusion, our results are in very good agreement with the features of the LDOS predicted by the field-theoretical investigations. 

\subsection{Extended Hubbard model at half-filling}\label{sec:ext}
\begin{figure}[t]
\includegraphics[width=0.495\textwidth]{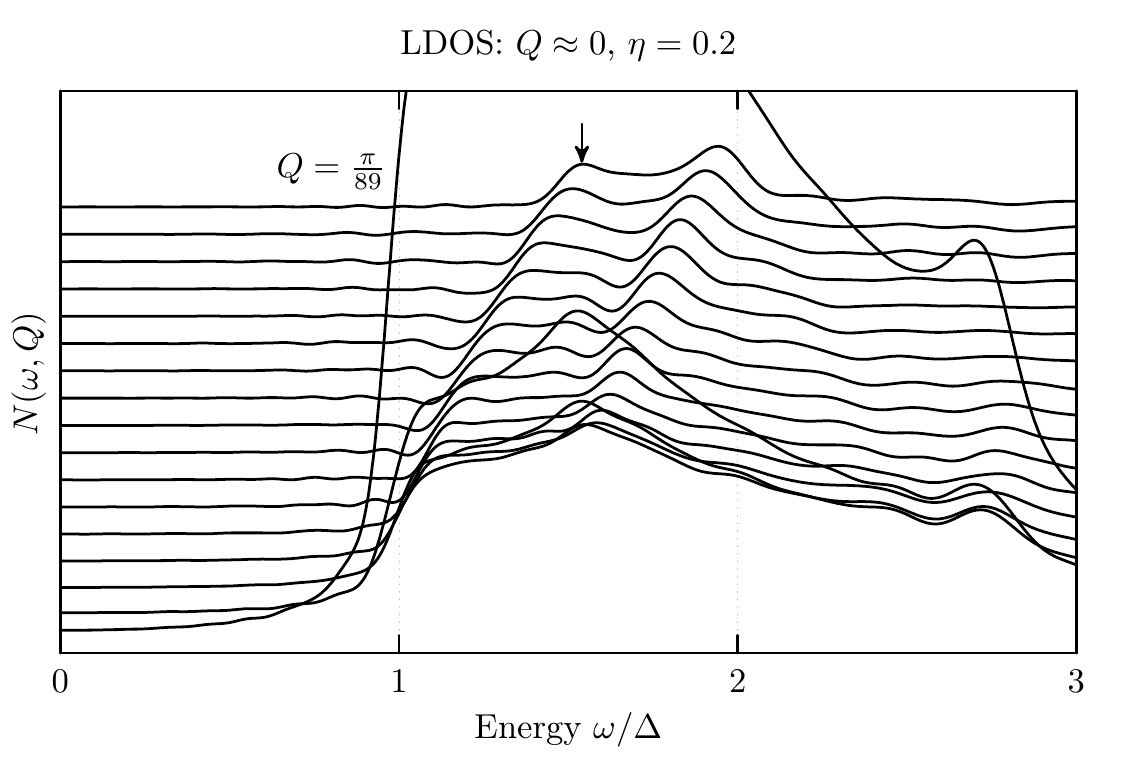}
\caption{$N(\omega,Q)$ for an extended Hubbard model with interaction $U=8$, $V=3$, $L=88$, $\eta=0.2$ and momenta $Q=\pi/89, 2\pi/89, \ldots, 17\pi/89$ (from bottom to top). The results are qualitatively similar to the ones for the standard Hubbard model shown in Fig.~\ref{fig:std_q0}, ie, we observe a Mott gap $\Delta$, a dispersing feature following \eqref{eq:holon} (indicated by the arrow) and another one at higher energies.}
\label{fig:hf_q0}
\end{figure}
\begin{figure}[b]
\includegraphics[width=0.495\textwidth]{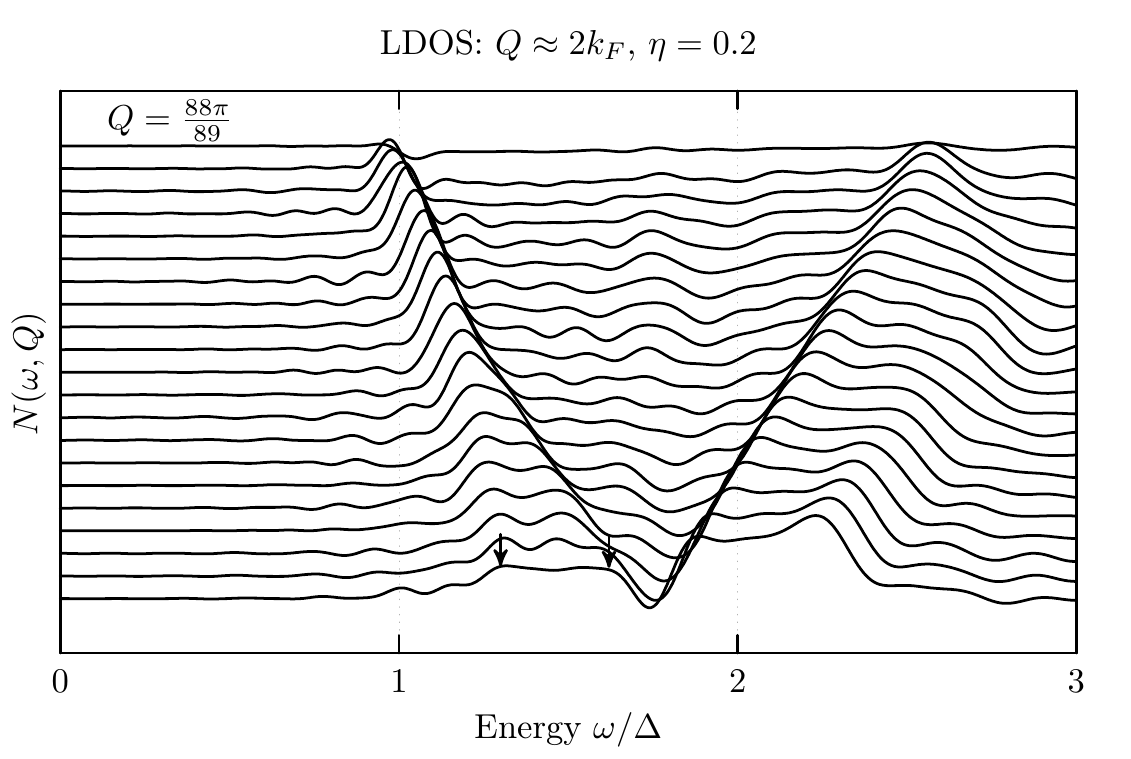}
\caption{$N(\omega,2k_\text{F}-q)$ for an extended Hubbard model in the vicinity of $Q=2k_\text{F}$ with $q=2k_\text{F}-Q=\pi/89, 2\pi/89, \ldots, 21\pi/89$ (from top to bottom).  All other parameters are as in Fig.~\ref{fig:hf_q0}. Similar to the standard Hubbard model, at low energies we observe two dispersing features at \eqref{eq:holon} and \eqref{eq:spinon} respectively.}
\label{fig:hf_q2kf}
\end{figure}

We have performed the analysis presented in the previous section for the extended Hubbard model (\ref{eq:ext}) at half-filling and $L=88$ lattice sites. Since the extended Hubbard model is not integrable, there exist no analytical results for the parameters $\Delta$, $v_\text{c}$ and $v_\text{s}$. Still, the field theory is expected to qualitatively describe the behaviour of the system in the low-energy limit. The LDOS for momenta in the vicinity of $Q=0$ and $Q=2k_\text{F}$ is shown in Figs.~\ref{fig:hf_q0} and~\ref{fig:hf_q2kf} respectively. In both plots we have renormalised the energy scale by the gap $\Delta\approx 2.1$ obtained from the data at $Q\approx 0$. 

At low energies the dispersing features are qualitatively identical to the ones seen for the standard Hubbard model, namely a propagating charge mode for $Q\approx 0$ and both a propagating charge and spin mode around $Q=2k_\text{F}$. The only difference is that the charge and spin velocities take the values $v_\text{c}\simeq 1.8\Delta\simeq 3.8$ and $v_\text{s}\simeq 0.35\Delta\simeq 0.7$ respectively, which were determined by comparison with the quasiparticle dispersions \eqref{eq:holon} and \eqref{eq:spinon}. The energy gap $\Delta$ and charge velocity $v_\text{c}$ for the two different momentum regimes agree well. We thus conclude that the low-energy sector is well described by the field theory. Furthermore, for small momenta we again observe a second charge mode which now seems to have the gap $\Delta_2\simeq 3\Delta/2$.

\section{Effect of a boundary potential}\label{sec:bbs}
Having analysed the LDOS in the presence of open boundary conditions, we now turn to the investigation of the effect of a boundary chemical potential. Specifically we consider the Hubbard model \eqref{eq:Hubbard} with a boundary potential at site $j=0$,
\begin{equation}
H_\text{bp}=H+\mu\sum_{\sigma} n_{j=0,\sigma}.
\label{eq:bcp}
\end{equation}
Using bosonisation such a boundary potential is translated into non-trivial boundary conditions for the bosonic degrees of freedom. In particular, certain boundary conditions give rise to the existence of boundary bound states in the gapped charge sector\cite{GhoshalZamolodchikov94} which manifest themselves\cite{SEJF08} in the LDOS as non-propagating features inside the Mott gap. The spectrum of the Hubbard chain with boundary potential \eqref{eq:bcp} has been investigated by Bed\"urftig and Frahm\cite{BeduerftigFrahm97} using the Bethe-ansatz solution. In particular it was found that a boundary bound state corresponding to a charge bound at the first site exists for $\mu<-1$. For even smaller boundary potentials, $\mu<-2u-\sqrt{1+4u^2}$, two electrons in a spin singlet get bound to the surface.

\begin{figure}[t]
\includegraphics[width=0.495\textwidth]{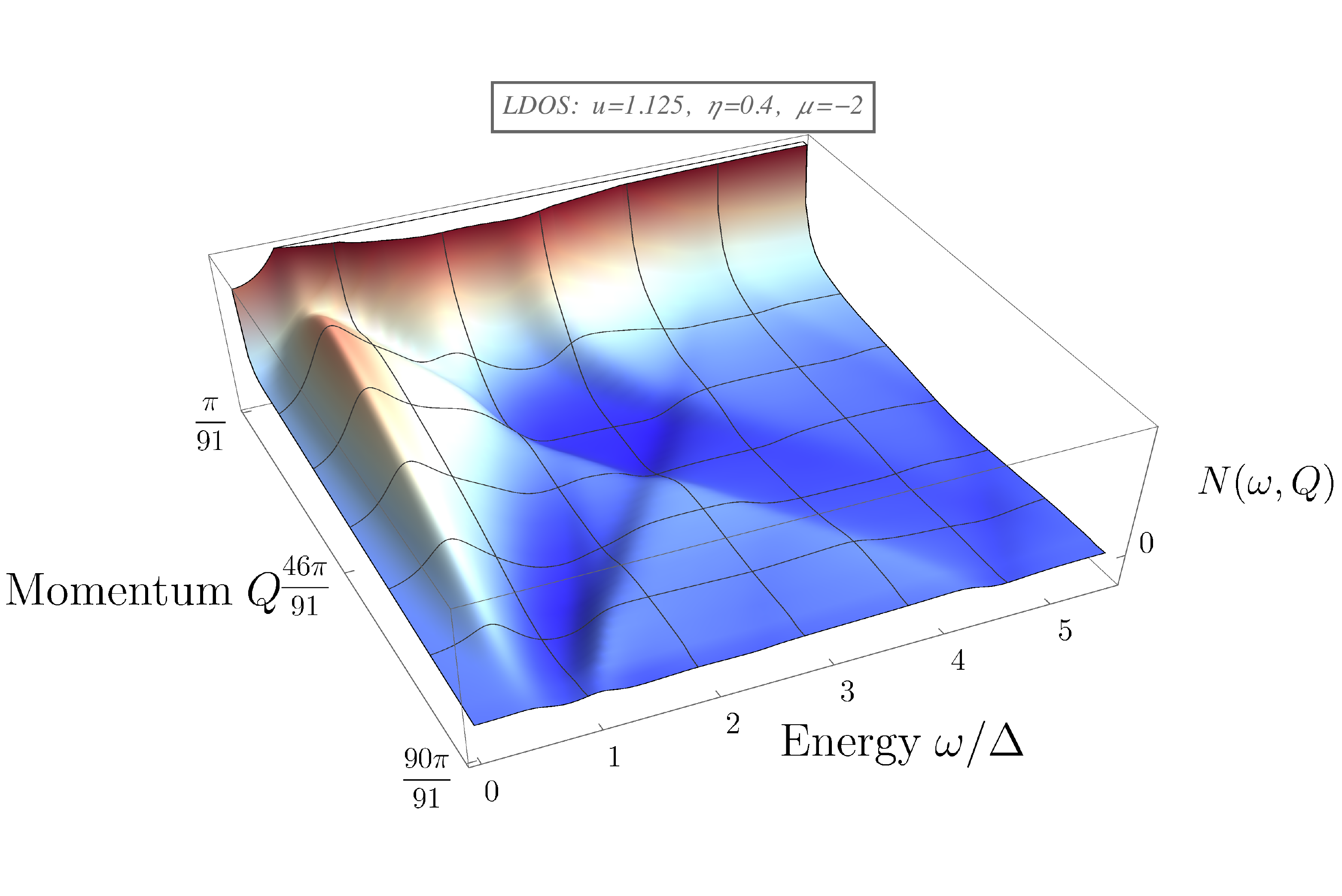}
\caption{$N(\omega,Q)$ for interaction $u=1.125$, boundary potential $\mu=-2$, $L=90$ lattice sites and broadening $\eta=0.4$. Besides the peak at $Q=0$ and the dispersing modes at $\omega\ge\Delta$ we observe a non-dispersing feature inside the energy gap at $\omega=E_{\text{bbs}}\approx \Delta/2$ which originates from the boundary bound state.}
\label{fig:bbs_q0}
\end{figure}
The Fourier transform of the LDOS in the presence of a boundary chemical potential is shown in Fig.~\ref{fig:bbs_q0}. Besides the peak at $Q=0$ due to the pinned charge-density wave and several dispersing modes above the Mott gap, we observe a clear, non-dispersing maximum inside the gap at $\omega=E_\text{bbs}\approx \Delta/2$, which is a manifestation of the boundary bound state in the LDOS. In the following we analyse this contribution in more detail by considering the LDOS $N(\omega,x)=-1/\pi\,\text{Im}\,G^\text{R}(\omega,x)$ close to the boundary. 

\begin{figure}[t]
\includegraphics[width=0.495\textwidth]{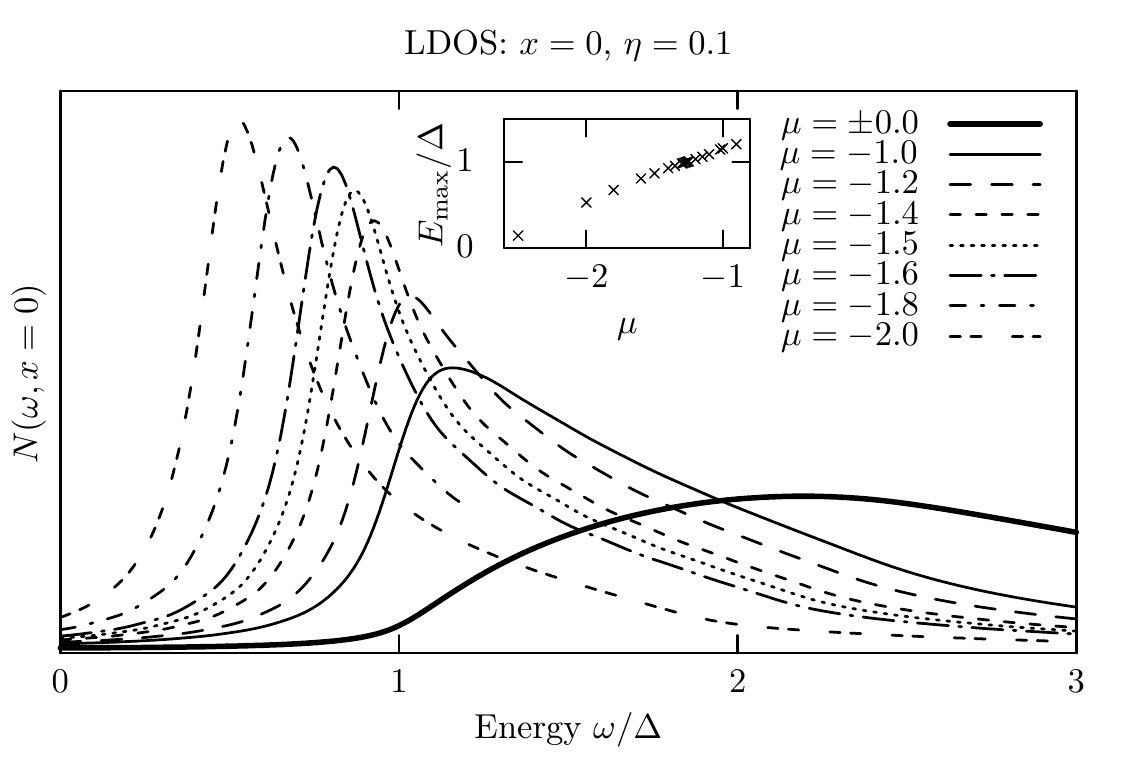}
\caption{LDOS at the boundary, $N(\omega,x=0)$, for various values of $\mu$ and broadening $\eta=0.1$. All other parameters are as in Fig.~\ref{fig:bbs_q0}. In the absence of a boundary potential (thick line) there is barely any spectral weight inside the energy gap. For $\mu<-1$ the spectral density inside the gap grows continuously but its maximum is still located above the gap. For $\mu\le -1.4$ the maximum is located inside the Mott gap, providing a clear manifestation of the boundary bound state. Inset: Position $E_{\text{max}}$ of the maximum of $N(\omega,x=0)$ as a function of the boundary potential $\mu$. We observe that a potential $\mu\leq -1.27$ is needed for $E_{\text{max}} < \Delta$.}
\label{fig:bbs_x0}
\end{figure}
\begin{figure}[t]
\includegraphics[width=0.495\textwidth]{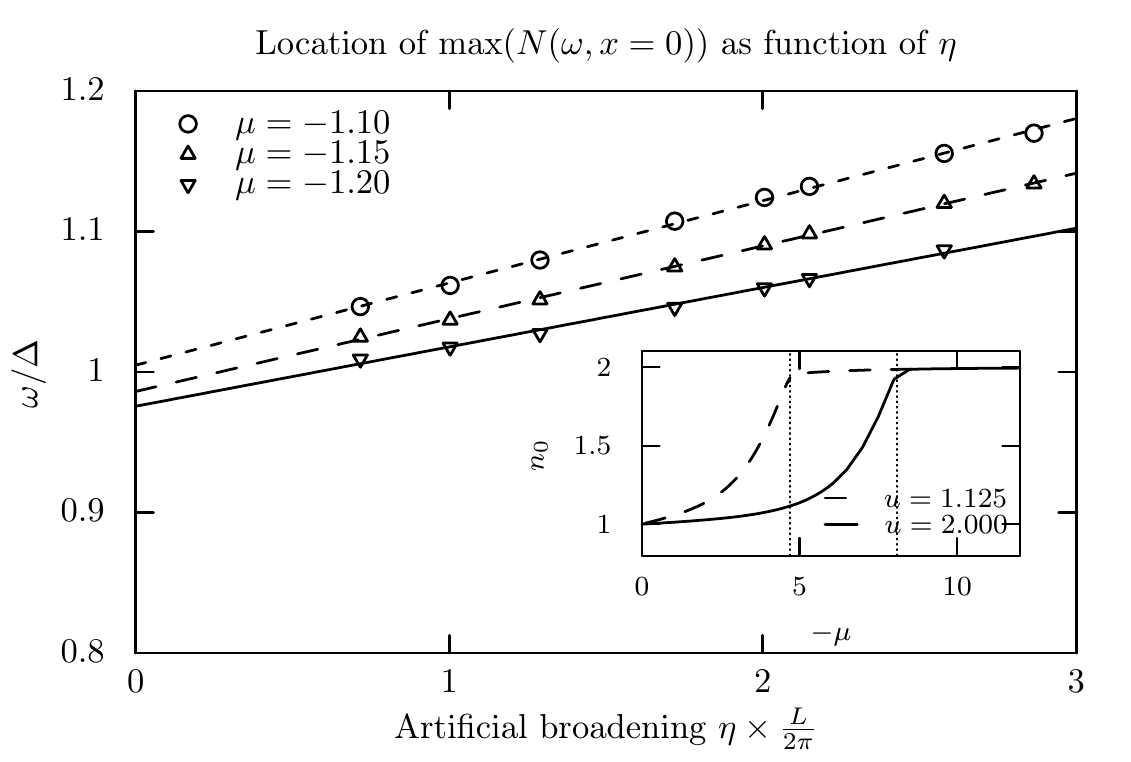}
\caption{Maximum of $N(\omega,x=0)$ as a function of the artificial broadening $\eta$ for $u=1.125$ and $L=90$. Extrapolating to $\eta=0$ (indicated by lines) we find that that the energy of the maximum lies within the Mott gap for $\mu\lesssim -1.15$. Inset: Electron density $n_0$ at the boundary showing very good agreement with the Bethe-ansatz result.\cite{BeduerftigFrahm97} The dotted vertical lines indicate the positions $\mu=-2u-\sqrt{1+4u^2}$ at which two electrons get bound to the boundary.}
\label{fig:bbs_mu}
\end{figure}

First we analyse the LDOS at the boundary site, $N(\omega,x=0)$, which is shown in Fig.~\ref{fig:bbs_x0} for several values of the boundary potential $\mu$ using an artificial broadening $\eta=0.1$. One can clearly see that the maximum of the LDOS is shifted towards lower energies for decreasing $\mu$. For $\mu\leq -1$ we find a considerable spectral density inside the Mott gap $\Delta$; for $\mu\lesssim-1.27$ the maximum of the LDOS is located inside the energy gap as well. From this we deduce that for $\mu\lesssim -1.27$ there exists a clear boundary bound state contribution to the LDOS. We attribute the deviation to the critical value $\mu=-1$ obtained from the Bethe ansatz\cite{BeduerftigFrahm97} to the finite system-size as well as the artificial broadening $\eta$ introduced in our numerical calculations. This is supported by the dependence of the energy of the maximum in the LDOS on the broadening presented in Fig.~\ref{fig:bbs_mu}, which shows that the energy of the maximum indeed decreases with decreasing $\eta$. Extrapolating the results to $\eta=0$ and keeping in mind the finite system size as well as the fact that for $\mu\to -1^-$ the contributions from the boundary bound state and the standard continuum at $\omega\ge\Delta$ start to significantly overlap, we conclude that our results are consistent with the Bethe-ansatz solution. This is further supported by the electron density at the boundary shown in the inset of Fig.~\ref{fig:bbs_mu}.
\begin{figure}[b]
\includegraphics[width=0.495\textwidth]{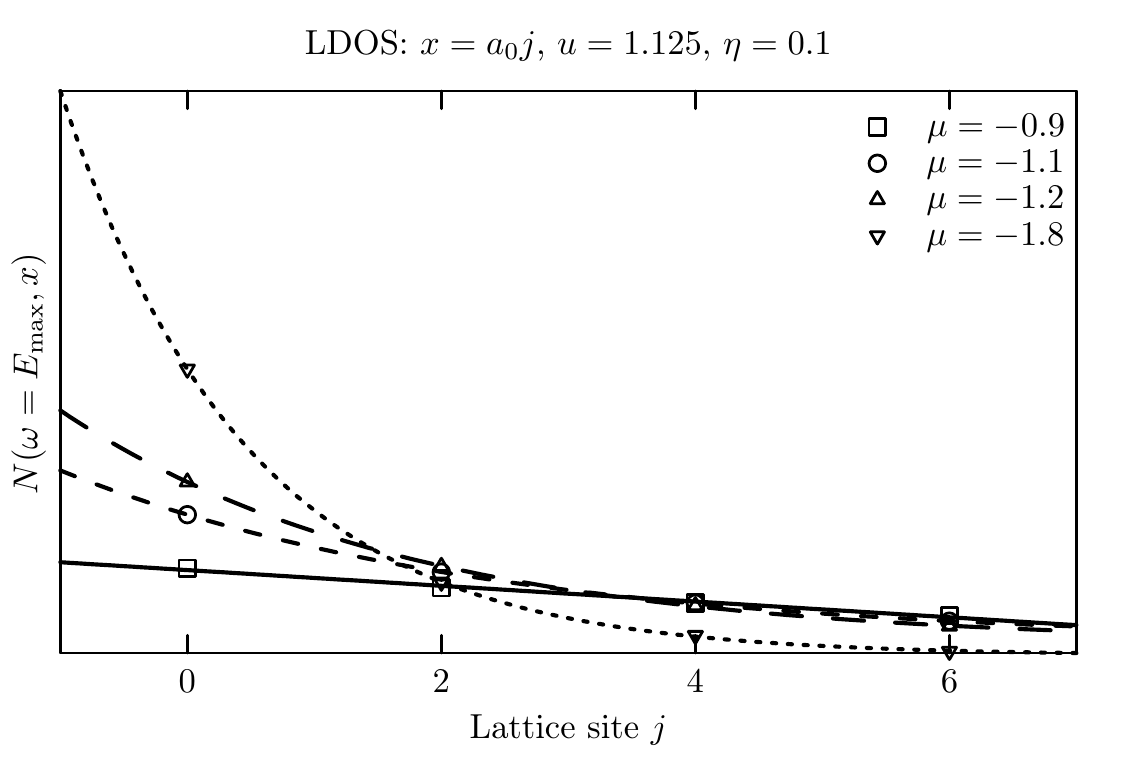}
\caption{Maximal value of the LDOS, $N(\omega=E_{\text{max}},ja_0)$, as a function of the distance to the boundary for $u=1.125$, $\eta=0.1$ and $L=90$. For decreasing $\mu$ we observe that the spectral weight gets more and more localised at the boundary.}
\label{fig:bbs_exp}
\end{figure}
Finally we consider the space dependence of the LDOS when going away from the boundary. As is shown in Fig.~\ref{fig:bbs_exp}, lowering the boundary potential leads to an increase of the LDOS at the boundary, consistent with the formation of a boundary bound state localised at $j=0$. However, the system size and energy resolution is not sufficient to unveil an exponential space dependence of the LDOS as predicted by the field-theory analysis,\cite{SEJF08} ie, $N(\omega,x)\propto \exp[-2x\sqrt{\Delta^2-E_\text{bbs}^2}/v_\text{c}]$.

\section{Conclusion}\label{sec:conclusion}
In this work we have performed a numerical study of the LDOS of one-dimensional Mott insulators with an open boundary. As microscopic realisations of the Mott insulator we have studied the (extended) Hubbard model at half filling. The results for the Fourier transform of the LDOS revealed the existence of the Mott gap as well as several gapped and gapless dispersing modes. These qualitative features were in perfect agreement with the results of field-theoretical calculations\cite{SEJF08} of the LDOS in the Mott insulator. Furthermore, we extracted quantitative values for the gap and velocities, which, in the case of the integrable Hubbard chain, were found to be in excellent agreement with the exact results.\cite{EsslerFrahmGoehmannKluemperKorepin05} Besides open boundary conditions we have also considered the effect of a boundary potential. For sufficiently strong potentials this results in the formation of a boundary bound state, which manifests itself in the LDOS as a non-dispersing feature inside the Mott gap. In summary, our results show that spin-charge separation and the formation of boundary bound states can be observed in the Fourier transform of the LDOS amenable to numerical simulations or scanning tunneling spectroscopy experiments.

\acknowledgements
We thank Fabian Essler, Holger Frahm, Martin Hohenadler and Volker Meden for useful comments and discussions. BS and DS were supported by the Foundation for Fundamental Research on Matter (FOM), which is part of the Netherlands Organisation for Scientific Research (NWO), under 14PR3168. PS was supported by DFG-SFB 1170 and the ERC starting grant TOPOLECTRICS (ERC-StG-336012). This work is part of the D-ITP consortium, a program of the Netherlands Organisation for Scientific Research (NWO) that is funded by the Dutch Ministry of Education, Culture and Science (OCW). The authors acknowledge support by the state of Baden-W\"urttemberg through bwHPC.



\begin{thebibliography}{10}
\bibitem{footnote1}
We note that the Refs.~\onlinecite{SEJF08} consider the case of a
  charge-density wave state where the gap appears in the spin sector. The
  results for the Mott insulator considered in our article are obtained by
  simply interchanging spin and charge sectors in
  Refs.~\onlinecite{SEJF08}.
\end{thebibliography}
\end{document}